\documentclass[12pt,letterpaper]{llncs}

\usepackage{adjustbox}
\usepackage{float}
\usepackage[T1]{fontenc}
\usepackage{graphicx}
\usepackage[utf8]{inputenc}
\usepackage{multicol}
\usepackage{multirow}
\usepackage{txfonts}
\usepackage[style=apa,sorting=nyt]{biblatex}
\begin{document}

\title{Hybrid Horizons: Policy for Post-Quantum Security}
\author{Anais Jaikissoon, Contact Information: anaisjai718@gmail.com}

\maketitle


\begin{abstract}
The Age of Artificial Intelligence is here. In 2025, there are few regulations governing artificial intelligence. While the expansion of artificial intelligence is going in a relatively good direction, there is a risk that it can be misused. Misuse of technology is nothing new and will continue to happen. The lack of regulation in artificial intelligence is necessary because it raises the question of how we can move forward without knowing what the limits are. While artificial intelligence dominates the technology industry, new technology is starting to emerge. Quantum cryptography is expected to replace classical cryptography; however, the transition from classical to quantum cryptography is expected to occur within the next 10 years. The ability to transition from classical to quantum cryptography requires hybrid cryptography. Hybrid cryptography can be used now; however, similar to artificial intelligence, there is no regulation or support for the regulatory infrastructure regarding hybrid machines. This paper will explore the regulatory gaps in hybrid cryptography. The paper will also offer solutions to fix the gaps and ensure the transition from classical to quantum cryptography is safely and effectively completed.

\end{abstract}

\section{Introduction}

It is normal for crypto systems to be created; in response, people will attempt to break them. An example of a broken cryptographic standard is the Data Encryption Standard (DES). After one standard has been broken, the next step is to utilize a system that has not yet been compromised to help protect data. DES was officially replaced by the Advanced Encryption Standard (AES), according to the National Institute of Standards and Technology (NIST). This cycle of creating and breaking is occurring once again. Current classical cryptography is being changed by a new system called Quantum Cryptography. This cryptography will revolutionize data protection in the next few years, shifting from classical cryptography to protect data at rest and in transit. Protecting data when it is being sent to someone else, also known as data in transit, is susceptible to cyberattacks. Data that is not being used, known as data at rest, is also susceptible to cyberattacks. The vulnerability associated with data handling underscores the importance of protecting it.

Quantum cryptography is designed to enhance data protection in transit and at rest by utilizing a new cryptographic method. However, this method is still being tested by the National Institute of Standards and Technology (NIST) to determine which of the four algorithms is quantum safe. The four algorithms being tested are CRYSTALS-Kyber, which is created for general encryption. CRYSTALS-Dilithium is the second one being tested, and it is designed to protect digital signatures. The third is SPHINCS+, similar to CRYSTALS-Dilithium in protecting digital signatures. The final algorithm, FALCON, is also used to protect digital signatures. Quantum cryptography is years away from being accessible to the public for several reasons, including determining the most effective algorithm for these computers, skepticism about transitioning from classical computers to quantum computers, and figuring out how to mitigate attacks against the data being held and transferred in these computers. There have been advancements in the private sector with these computers; IBM and the Open Quantum Safe project have developed quantum cryptography software that offers multiple resources, allowing users to test quantum encryption. Although the government expects everyone to shift from classical to quantum cryptography, the transition will not be smooth. There are currently few resources to help start the transition, and the public, while currently focusing on artificial intelligence, does not know what quantum cryptography is. Those outside the quantum and cybersecurity industry are often unaware of the difference between classical and quantum cryptography. 

This paper will examine current regulations, laws, standards, and frameworks regarding classical cryptography to help build a basic understanding of what is required to protect data in the classical sector. Then, the paper will explain what hybrid cryptography is. Following it will give recommendations to transition from classical to hybrid cryptography. The paper examines any current regulations, standards, and frameworks regarding hybrid machines. After examining hybrid regulations and standards, a brief outline of what hybrid regulations should be. The paper will then identify the risks associated with implementing quantum safe cryptography in hybrid machines and finally offer a new timeline of when the transition from classical to hybrid to quantum should occur.

\vspace{1\baselineskip}
\section{Classical Cryptography}

Cryptography has been around for centuries, but cryptography regulation has only existed since 1977 (NIST, 1977). Cryptography can be traced back to the Spartans \cite{subramani2023review}.  The most famous example of a cryptography system is the Caesar cipher, in which the recipient needs to know the key, and each letter is shifted based on the key. In 1976, the Diffie-Hellman key exchange was published, followed by RSA in 1977. In 1977, the Data Encryption Standard (DES) became Federal Information Processing Standards (FIPS) 46 (NIST, 1977). Now there will be quantum cryptography and post-quantum cryptography. Quantum cryptography is based on the natural and unchanging laws of quantum mechanics \cite{the2025history}. These advances in technology have allowed for better security \cite{schneider2024history}. In 1976, the Diffie-Hellman key exchange was published, followed by the publication of RSA in 1977. In 1977, the Data Encryption Standard (DES) became Federal Information Processing Standards (FIPS) 46 (NIST, 1977). Now there will be quantum cryptography and post-quantum cryptography. Quantum cryptography is based on the natural and unchanging laws of quantum mechanics  \cite{nist2019fips140}. These advancements in technology have allowed for better security. 

 Better security means users can ensure their data is safe, whether it is in transit or at rest. With any technology, questions of misuse are being raised. Some questions include: What if the organization that holds the data misuses it? What if the organization made a mistake that caused harm to the user or the data? Regulations, laws, standards, and frameworks have been proven to help organizations comply with the law and protect victims from exploitation. However, organizations can use the latest technology as they see fit if there are no regulations, laws, standards, or frameworks. These safeguards can also help organizations understand the latest technology that is being released and how to migrate to it. Without proper guidance from the government and organizations like the National Institute of Standards and Technology (NIST), companies and other organizations will struggle to migrate to quantum cryptography. The ability of a user to understand the latest technology comes first from understanding the history of the technology from which it evolved.

\vspace{1\baselineskip}
\subsection{Review of current Standards, Frameworks, and Guidelines for Classical Cryptography}

NIST is currently the go-to for understanding how to navigate new technologies such as artificial intelligence (AI). NIST is also a non-governing organization that aids others to help protect their data. An example of NIST helping users is by releasing multiple regulations, laws, standards, and frameworks for classical cryptography. NIST has released multiple standards for classical cryptography, including Federal Information Processing Standards (FIPS) 46, Federal Information Processing Standards (FIPS) 140-3, and Special Publication (SP) 800-57. FIPS 46 was originally published on January 15, 1977. This standard was implemented when organizations stated that there was ``cryptographic protection" being used for sensitive data. The standard explains the importance of protecting data that is in transit or at rest. FIPS 46 was used as a guide for users on how to change data to a cryptographic cipher. It also specifies how to decrypt the encrypted data. The cryptographic cipher was the Data Encryption Standard (DES). Its purpose was to protect data, as most cryptographic ciphers are supposed to do. The standard was released a year after the Diffie-Hellman key exchange was published, demonstrating the government's ability to produce guidelines concerning data protection.

Forty-two years after FIPS 46 was published, FIPS 140-3, also known as Security Requirements for Cryptographic Modules, was released. FIPS 140-3 was published on March 22, 2019. The primary purpose of ensuring that the hardware, software, and firmware that use encryption are designed to be \cite{iso2025iso19790}. This standard specifies the security requirements for cryptographic modules that are used to protect sensitive data. It sets the design and architectural design for cryptographic modules. This includes using algorithms approved by NIST, such as AES, SHA-2, and RSA. The standard also states that there must be design features \cite{nist2020sp80057}, including access control and destruction of information that is no longer needed. The standards also emphasize the need for physical and logical controls when protecting a secure area. Although physical and logical controls are not protected under cryptography, it is still vital for data protection. 
The equivalent of FIPS 140-3 on an international level is ISO 19790 \cite{hulsingHashBased}; both are designed to focus on physical and logical security. ISO 19790 and FIPS 140-3 are used for the development and security of cryptographic modules. There must be design features that help protect sensitive data, including access control and destruction of information no longer needed. 

NIST released SP 800-57 in May 2020, titled ``Recommendation for Key Management". This SP gives guidance on how to manage cryptographic keys throughout their life cycle. The life cycle includes creation, use, storage, distribution, and destruction. It first explains the reason for cryptography \cite{barker2020keyderivation} ,provides  support for the cybersecurity functions of Confidentiality, Integrity, and Availability (CIA). It then describes the importance of non-repudiation and its relationship to cryptography, specifically digital signatures, which ensure that the person with the signature cannot deny having the key. The publication provides detailed information on approved algorithms that have been thoroughly tested and are deemed adequate for security purposes. These algorithms are cryptographic hash functions, symmetric-key algorithms, and asymmetric-key algorithms. The document then gives guidance on the key management of Confidentiality, Integrity and Availability (CIA). It then describes the importance of non-repudiation and its relationship to cryptography, specifically digital signatures, which ensure that the person with the signature cannot deny having the key. 

 SP 800-57 was published a year after FIPS 140-3, and similar to the documents previously mentioned, emphasizes a need to protect data using classical cryptographic ciphers. Now, governments are shifting their focus to quantum cryptography, which is intended to protect data from attacks executed on quantum computers. However, the quantum computer has already been created. IBM has its own quantum computer \cite{ibmQuantum}. There have been reports that our adversaries, such as China, have a quantum computer. How do other governments protect themselves from an attack executed on a quantum computer? Hybrid cryptography is a temporary solution that will help mitigate the impact of these attacks. 

\section{Recommendations for Hybrid Machines}

\subsection{What is Hybrid Cryptography?}

Hybrid cryptography is the use of classical and quantum cryptography. A classical computer cannot use quantum cryptography to protect the data because the classical computer does not have enough resources, and quantum cryptography is built on the laws of physics, unlike classical which is built on mathematics. Quantum cryptography requires a large amount of space that a classical computer cannot simply handle. Because quantum computing is based on superposition — the equivalent of an ‘and’ statement, where multiple outcomes can exist simultaneously — classical computers cannot handle this type of computation. Classical computers are built on ``or" statements that entail one outcome or the other outcome, and both outcomes cannot happen at the same time. Quantum computers are built on ``and" statements so situation A and B can happen at the same time \cite{chamberlain2025blackgirls}. Due to the ``and" statement occurring in quantum computers, it makes it easier to execute attacks at a faster rate. However, as quantum computers become more popular, vulnerabilities to these attacks are beginning to appear. Hybrid cryptography can help mitigate the risk of attacks. 

 Hybrid cryptography entails the use of classical and quantum cryptography in a machine. For example, using a quantum-resistant signature such as the Leighton-Micali Signature (LMS) or the eXtended Merkle Signature Scheme (XMSS) will reduce the chance of a quantum attack because these two are supposed to be strong enough to handle a quantum attack. The user does not need both LMS and XMSS on the system because one signature alone takes up space on the machine. The need for a classical cryptography such as Secure Hash Algorithm (SHA) 256, which is considered secure against attacks currently, is to help reduce the chances of an attack from a classical computer. Artificial intelligence makes it easier to make, execute, and purchase attacks nowadays, so there is still a need for classical cryptography to protect data.

Regarding the physical aspects of the hybrid computer to be able to compute both quantum-resistant and classical cryptography, there needs to be multiple Central Processing Units (CPU) with at least six cores to aid with quantum computation. The amount of RAM will need to be more than 256GB because even doing basic office work such as web browsing or researching requires the 256GB. If an organization decides to use the hybrid machine for multi-use such as web browsing and protecting data with the hybrid cryptography, more RAM will be needed. One way to mitigate this if obtaining more RAM is not possible is to get a Solid-State Drive (SSD) to help store the data from web browsing.

\subsection{Recommendations for Transition from Classical to Quantum Cryptography using Hybrid Machines}

Organizations must first use hybrid cryptography if they want a smooth transition from classical cryptography to quantum cryptography. Using quantum-resistant cryptography changes the environment of the classical computer because now there is another way of encrypting and decrypting data. Users who desire a hybrid computer must first test quantum-resistant cryptography in an isolated environment with fake data. There are multiple ways to test quantum cryptography on a hybrid computer; one approach is to utilize the Open Quantum Safe \cite{oqsFAQ} (OQS) project. This can be achieved on a classical computer that utilizes quantum cryptography to encrypt data. The liboqs is an open-source library created by OQS containing quantum-resistant cryptography, which, in theory, allows data in transit to be resistant to quantum attacks. OQS states that the project utilizes a Module-Lattice-Based Key-Encapsulation Mechanism (ML-KEM) and a Module-Lattice-Based Digital Signature Algorithm (ML-DSA). NIST has approved ML-KEM and ML-DSA as quantum-resistant cryptographic methods. The implementation process is relatively simple and can be done using Linux, macOS, and Windows, which require more than eight gigabytes of RAM. However, OQS has not been NIST certified, so it cannot be guaranteed that the data will be protected from quantum attacks. OQS is working to mitigate risks and ultimately make it safe to handle sensitive data.

This testing phase is essential: A user who is stress testing the environment can mimic an attack from a quantum computer; with fake data in the computer, no real data will be affected. The next tool that can help mimic an attack that can also be used on a classical computer is the IBM tool, Qiskit, which helps classical computers mimic a quantum computer using quantum cryptography. Qiskit also contains Qiskit Aer, which can simulate a real quantum computer; in turn, it can mimic an attack from a quantum computer (IBM, n.d.). Within Qiskit, a feature named Benchpress allows users to perform 1,000 different tests to benchmark their quantum cryptography \cite{qiskitBenchpress2024}. Once OQS and Qiskit tools have been successfully implemented in the classical computer, it becomes a hybrid computer due to its capabilities to use classical cryptography and quantum-resistant cryptography.

\subsection{Current Regulations for Hybrid Cryptography Machines}

Currently, some standards enable machines to operate using hybrid cryptography. However, some standards have been released that could be used for hybrid cryptography regulation. One is NIST SP 800-208, ``Recommendation for Stateful Hash-Based Signature Schemes."  This SP specifies two algorithms that can help generate a digital signature based on hash-based signatures. The first is the Leighton-Micali Signature (LMS), and the second is the eXtended Merkle Signature Scheme (XMSS). LMS and XMSS are considered quantum-resistant because they are hash-based signatures (NIST, 2024). This SP enables the use of LMS and XMSS with various hash functions, including SHA-256 and SHA-256/256. Both quantum-resistant cryptography (LMS or XMSS) and classical cryptography (SHA-256) enable this combination of cryptographic systems in a hybrid machine. Classical cryptography can protect against current threats, while implementing quantum-resistant cryptography allows protection from future threats. SP 800-208 is not explicitly made for hybrid cryptography because there is no mention of hybrid cryptography. However, implementing quantum-resistant and classical cryptography can help facilitate the transition. LMS and XMSS are quantum-resistant because they are hash-based signatures. This SP allows the use of LMS and XMSS with different hash functions such as SHA-256 and SHA256/256. Using both quantum-resistant cryptography (LMS or XMSS) and classical cryptography (SHA-256) allows for this mash of cryptographic systems in a hybrid machine.

NIST has published SP 800-56C Rev. 2, ``Recommendation for Key-Derivation Methods in Key-Establishment Schemes." SP 800-56C Rev.2 explains techniques for key derivation of a shared secret. Once two parties agree on a secret, SP 800-56C Rev. 2 provides the steps to help the two parties transform the secret into cryptographic keys. The steps are key agreement, shared secret, key derivation, and cryptographic keys. The publication allows for two types of key derivation: the first one being One-Key Derivation, which uses a One-Step Key Derivation Function (KDF) to derive the key from the shared secret. The cryptographic method can be a hash function or a Message Authentication Code (MAC) function to help obtain the key. The second method explained is Two-Step Key Derivation, which extracts the Key Derivation Key (KDK) from the shared secret using a key extraction function and then expands the key required to acquire it. One-Key Derivation can be used for simple systems, while Two-Step Derivation can be used for stronger cryptographic systems. However, similarly to SP 800-208, SP 800-56C Rev.2 never explicitly addresses hybrid cryptography. Both methods are suitable for hybrid cryptography because they allow users to use classical and quantum cryptography in their systems.

A document mentioning hybrid cryptography originates from the European Telecommunications Standards Institute (ETSI). This guideline is titled Migration Strategies and Recommendations to Quantum-Safe Schemes, also known as ETSI GR QSC V1.1.1. This document gives strategic guidance for migrating to quantum cryptography from classical cryptography. The document emphasizes that quantum cryptography could break current classical cryptography, such as SHA-256. It outlines the phases organizations should initiate now. The first step is to complete an inventory of all cryptographic assets, assess the risk, and plan for migration, which could involve either entirely switching to quantum or transitioning to hybrid cryptography, and then switching to quantum. The next step is to deploy and test the quantum safe cryptography. The final step is to monitor the systems \cite{etsi2018qkd007}. The document states that using hybrid cryptography is a way to ensure that systems are compatible with quantum technology and can help minimize risks now.

\subsection{What Regulation for Hybrid Cryptography Should Look Like}

While users are preparing to implement or are currently implementing quantum cryptography, NIST should release multiple FIPS and SP on hybrid cryptography. Regulation of cryptography is necessary to help mitigate the misuse of quantum cryptography and hybrid cryptography. Without regulation, there will be misuse not only by users but also by the organizations that use this technology. There have been numerous cases where organizations have misused artificial intelligence, and due to the regulations in place, victims have been able to seek justice. The misuse of quantum computing and cryptography is inevitable, but policymakers and lawmakers should anticipate and address this issue now. The first step in doing so is to establish formal regulations for hybrid machines, which include creating multiple FIPS and SP standards. The FIPS should be created as a means of ensuring the federal government protects sensitive data using quantum cryptography. The FIPS for hybrid cryptography should be based on ETSI GR QSC V1.1.1, NIST 800-208, FIPS 140-3, and SP 800-56C Rev.2. 

 SP in hybrid cryptography would include an introduction, scope, and purpose. The scope and purpose would encompass recommendations for various cryptographic methods that can be utilized in hybrid cryptography for key derivation.  Similarly to SP 800-56C Rev. 2, the approved techniques for encrypting and decrypting the shared secret will also be mentioned. The life cycle of the key derivative will be performed using One-Key Derivation, and a list of systems on which it can be used will be provided. This includes any system that has data classified as low \cite{stine2008mapping} systems that handle data that are classified as anything above moderate. When creating this, a random bit generator is needed, and the generator has to be approved by NIST. Based on the guidelines given by NIST, classification levels have been released for law enforcement, medical, defense, and administrative organizations regarding how each organization should categorize its data (NIST, 2004). 

Quantum cryptography can be used to create digital signatures; this requires an SP for hybrid cryptography regarding digital signatures. This document will follow the same format as NIST SP 800-208. The SP will highlight the importance of two algorithms that can help generate digital signatures based on hash-based signatures. It will include two algorithms, one classical and one quantum. The classical algorithm will consist of pairing SHA-256 or SHA-256/256 to generate a digital signature, because quantum computers have not yet broken it. SHA-256 is the current recommendation by NIST for classical cryptography. The second cryptographic method would be quantum-resistant, providing for a Module-Lattice-Based Key-Encapsulation Mechanism (ML-KEM) or Module-Lattice-Based Digital Signature Algorithm (ML-DSA). For example, SHA-256 would be paired with LMS, which is a hash-based signature, to produce a digital signature. 

Although producing SP and FIPS are good starting points, there still need to be laws, frameworks, and an emphasis on compliance in quantum cryptography. There have been advances in the sector, but frameworks, standards, and regulations remain necessary to mitigate the risk of misuse within the United States. Currently, there is little documentation available regarding hybrid machines. The United States is spending billions of dollars on quantum computers and cryptography. However, it overlooks the risks associated with transitioning from classical computers to quantum computers without proper training or guidance. The United States has requested $\$$998 million for quantum computing \cite{subcommittee2024nqi} , which pales in comparison to IBM’s $\$$150 million investment in quantum \cite{ibm2025quantumstarling} The regulations, frameworks and laws that are released should allow for more funding to be allocated to quantum and hybrid cryptography. Without proper regulations, frameworks, and laws, there is a lack of resources allocated to hybrid cryptography, which can lead to numerous threats that arise from various situations. 

\subsection{Risk Associated with Hybrid Cryptography}

Many risk factors can arise when implementing a new system; however, there is a greater emphasis on mistakes because the field of quantum computing and cryptography is unique. The field of quantum computing is based on qubits and differs significantly from the classical principles of 0 or 1 \cite{chamberlain2025blackgirls}. Due to the lack of standards, incorrect implementation can occur because organizations may not know how to implement quantum cryptography into their systems. Incorrect implementation can lead to sensitive data being at increased risk of Man-in-the-Middle attacks or data breaches.  The ability to do hybrid or quantum cryptography requires a large amount of space. Organizations may not have the resources to invest in quantum cryptography at present, but with hybrid cryptography, they do not have to devote significant time or resources to this technology. Unfortunately, this can lead to accessibility problems when users or organizations with limited budgets cannot transition. This can also lead to more data breaches from these organizations that cannot afford to implement it. The innovation of quantum cryptography is poised to thrive and transform the world; however, its misuse can also occur. The misuse of hybrid cryptography can occur because there are no proper regulations, frameworks, or laws in place. Given the history of misuse of technology, it is likely that attackers could exploit hybrid cryptography itself. 

 There needs to be a clear distinction between what is allowed and what is illegal when using hybrid and quantum cryptography. Below are two risk matrices. The first one is the risks associated with using hybrid cryptography, and the second one is the risk associated with an organization choosing to stay with classical cryptography until quantum cryptography is readily available to the public. The likelihood and threats will be based on a scale of one to five, with one being the least likely scenario and least impact, and five being the most likely scenario and most impact. The numbers will be based on the frequency of previous attacks and the impact. The risk score is the calculation of likelihood times impact; the higher the score ,the more catastrophic the threat is.

\begin{table}[H]
\begin{adjustbox}{max width=\textwidth}
\begin{tabular}{p{4.13cm}p{4.13cm}p{4.13cm}p{4.13cm}}
\hline
\multicolumn{1}{|p{4.13cm}}{\textbf{Threat}} & 
\multicolumn{1}{|p{4.13cm}}{\textbf{Likelihood}} & 
\multicolumn{1}{|p{4.13cm}}{\textbf{Impact}} & 
\multicolumn{1}{|p{4.13cm}|}{\textbf{Risk Score}} \\ 
\hline
\multicolumn{1}{|p{4.13cm}}{Data Breach} & 
\multicolumn{1}{|p{4.13cm}}{3} & 
\multicolumn{1}{|p{4.13cm}}{5} & 
\multicolumn{1}{|p{4.13cm}|}{15} \\ 
\hline
\multicolumn{1}{|p{4.13cm}}{Harvest Now, Decrypt Later (HNDL)} & 
\multicolumn{1}{|p{4.13cm}}{2} & 
\multicolumn{1}{|p{4.13cm}}{5} & 
\multicolumn{1}{|p{4.13cm}|}{10} \\ 
\hline
\multicolumn{1}{|p{4.13cm}}{Ransomware attacks} & 
\multicolumn{1}{|p{4.13cm}}{2} & 
\multicolumn{1}{|p{4.13cm}}{5} & 
\multicolumn{1}{|p{4.13cm}|}{10} \\ 
\hline

\end{tabular}
\end{adjustbox}
\caption{Using Hybrid Cryptography}
\end{table}

Table 1 shows three different risks that an organization would face if it migrated to hybrid cryptography. Data breaches can be caused by an insecure implementation. Lack of government guidelines can prevent the organization from properly installing quantum cryptography. Inadequate implementation can lead to a data breach because the information is not correctly installed. However, due to the classical cryptography still in the system, there is a chance that the information will not be breached. The likelihood of a data breach is three because data breaches are becoming more common. Organizations with relaxed cybersecurity are more likely to become victims of a data breach \cite{osanoPrivacyBreach}. The impact of data breach would be five because this can cause numerous situations in which the organization is losing money. When a data breach occurs, an organization should offer free data monitoring to affected individuals. If the organization does or does not offer data monitoring, its reputation is tainted and no longer considered trustworthy. In return, this can cause a loss of customers and eventually money. The risk score that is over 15 should be mitigated sooner rather than later.

The second risk is Harvest Now, Decrypt Later (HNDL), and the likelihood of this is currently low due to the limited availability of quantum computers and cryptography. However, the likelihood is expected to increase as more quantum computers are available. The impact of this attack is high because attackers are gaining information and trying to decrypt it. This attack allows attackers to harvest the information. If the attacker cannot immediately break the encryption, then with the help of quantum computers, the attacker can try different methods at once to try and break the encryption. The HNDL risk score is 10, meaning that if an organization were to be faced with this attack, it would be significant.

The last risk is a ransomware attack similar to the HNDL. The likelihood of this is low. While ransomware attacks are increasing, a successful ransomware attack on a hybrid machine is low. The mix of quantum-resistant and classical cryptography allows for defensive encryption on both the quantum and classical fronts. Hybrid cryptography would be one defensive measure against the attack, and there would be an Intrusion Detection System (IDS) and an Intrusion Prevention System (IPS) along with other defensive measures. However, the impact of a ransomware attack would be significant. Due to the low likelihood of an HNDL, it puts the event in the significant category; however, the likelihood of these attacks will become more frequent. This would change the score and become a catastrophic event.

The use of hybrid cryptography is the best way to not only defend against these risks, but also to start the process of migrating to quantum cryptography. If an organization does not start the process now, there is a risk matrix that explains the likelihood and impact of the same risk, but only using classical cryptography against hybrid or full quantum computers. 

\begin{table}
\begin{adjustbox}{max width=\textwidth}
\begin{tabular}{p{4.13cm}p{4.13cm}p{4.13cm}p{4.13cm}}
\hline
\multicolumn{1}{|p{4.13cm}}{\textbf{Threat}} & 
\multicolumn{1}{|p{4.13cm}}{\textbf{Likelihood}} & 
\multicolumn{1}{|p{4.13cm}}{\textbf{Impact}} & 
\multicolumn{1}{|p{4.13cm}|}{\textbf{Risk Score}} \\ 
\hline
\multicolumn{1}{|p{4.13cm}}{Data Breach} & 
\multicolumn{1}{|p{4.13cm}}{4} & 
\multicolumn{1}{|p{4.13cm}}{5} & 
\multicolumn{1}{|p{4.13cm}|}{20} \\ 
\hline
\multicolumn{1}{|p{4.13cm}}{Harvest Now, Decrypt Later (HNDL)} & 
\multicolumn{1}{|p{4.13cm}}{2} & 
\multicolumn{1}{|p{4.13cm}}{5} & 
\multicolumn{1}{|p{4.13cm}|}{10} \\ 
\hline
\multicolumn{1}{|p{4.13cm}}{Ransomware attacks} & 
\multicolumn{1}{|p{4.13cm}}{4} & 
\multicolumn{1}{|p{4.13cm}}{5} & 
\multicolumn{1}{|p{4.13cm}|}{20} \\ 
\hline
\end{tabular}
\end{adjustbox}
\caption{Using Classical Cryptography}
\end{table}

According to one study, the number of data breaches has increased from 447 in 2012 to 3,200 in 2023 \cite{varonisDataBreachStats} , and the number is expected to rise. This puts the likelihood of a successful data breach at four because different factors can cause a breach, such as internal human error or bad actors. Classical cryptography cannot protect the data against an attack from a hybrid or quantum computer. If an organization has an Intrusion Detection System (IDS) or an Intrusion Prevention System (IPS), it may trigger the systems but based on previous data breaches the intrusion can go months without notice by the cybersecurity team. Any data breach will have an impact on the server and is rated five. This causes the risk score to be a catastrophic event.

HNDL attacks are on the rise because it uses quantum cryptography to break the encryption, and there is no need for a quantum computer; the attacker can use a hybrid computer. Currently, the likelihood is two because of the limited availability of knowledge and resources for quantum cryptography. However, in a few years the number will increase to four. The impact is still a five because, similar to a data breach, any breach will cause damage to the organization’s money and reputation. This puts the risk score at 10; however, it is expected to rise to 20 and become catastrophic.

The final risk is ransomware attacks; without quantum cryptography, these attacks increased 50$\%$ at the end of 2024 \cite{nortonRansomwareStats}. The number of attacks is going to increase as quantum cryptography becomes more available. In 2023, 66$\%$ organizations reported being victims of a ransomware attack \cite{sophos2024ransomware} . This puts the probability at four because the event is happening frequently. The impact is a five due to the nature of the attack, does the organization pay the ransom, or try to fight back? Both options would cause a loss of money and a significant amount of time would be spent trying to repair the situation. The score for this is 20 and is the catastrophic section because of the high likelihood and profound impact of a ransomware attack. 

\subsection{Timeline for Hybrid Cryptography}

Currently, NIST has a timeline for quantum computing that started in 2016 and is expected to end in 2035. From 2025 to 2035, there is an integration and application phase that is supposed to help the transition. NIST released an Internal Report (IR) 8547, which explains the timeline for migration from classical to quantum cryptography.

Below is a timeline outlining the transition from classical to hybrid to quantum cryptography. The first box indicates that, in 2026, NIST should release multiple FIPS and SP in hybrid cryptography. According to the current timeline provided by NIST, there are no plans to release a FIPS for hybrid cryptography in 2026. If NIST releases a FIPS in 2026, it can better support the original timeline provided by NIST. 2026 gives organizations enough time from 2026 to 2030 to transition to quantum cryptography. However, the timeline below gives organizations less time to transition from classical to hybrid cryptography. This is because the transition from classical to hybrid cryptography would be easier for organizations due to their previous knowledge of classical cryptography and the use of the same machines to help execute quantum cryptography. The second box states that in 2027, NIST needs to release frameworks so organizations can become compliant by 2029 in hybrid machines and cryptography.

\begin{figure}[H]
\includegraphics[width=14.33cm,height=8.08cm]{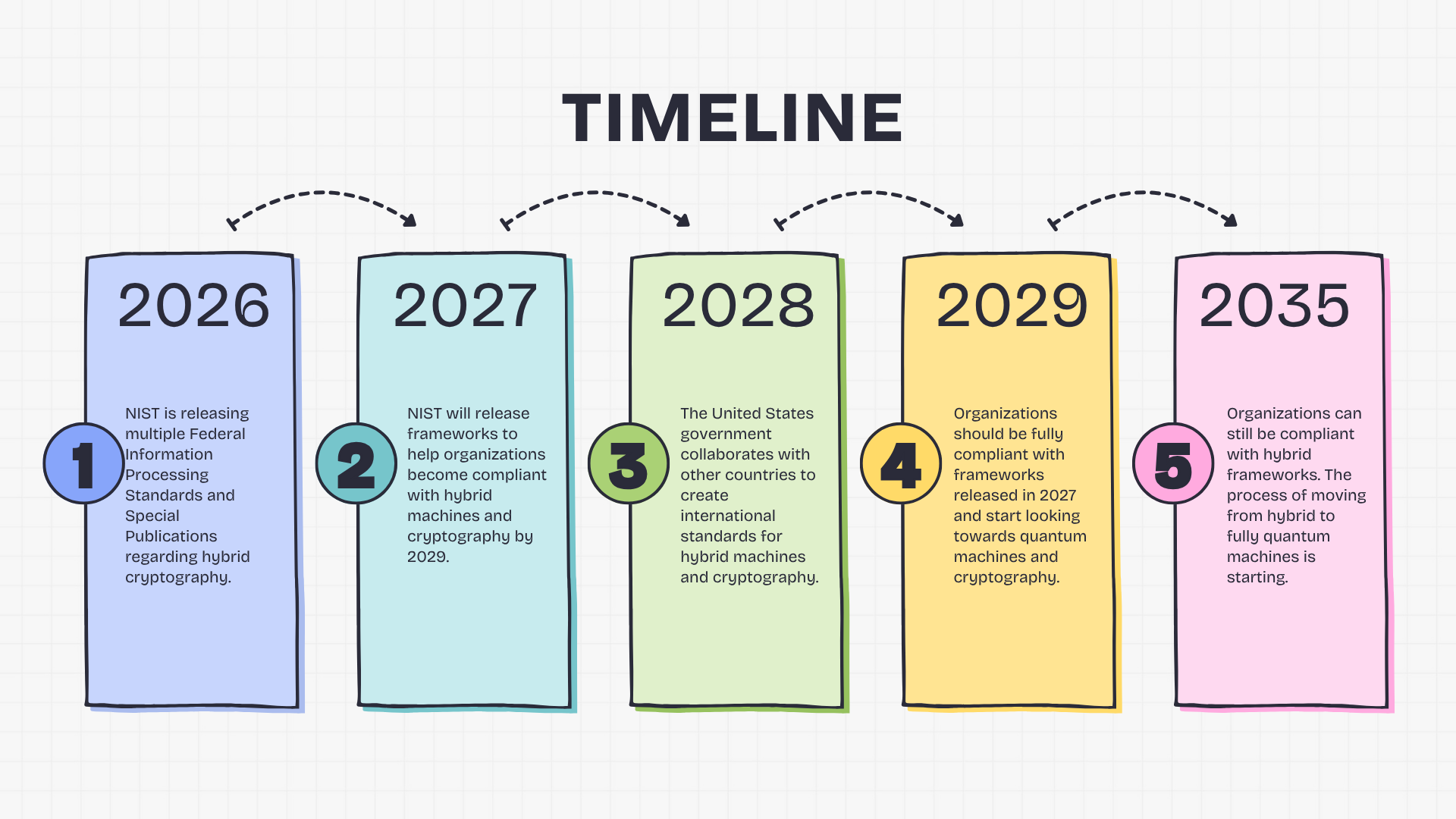}
\caption{PQC Migration}
\label{fig:pqcmain}
\end{figure}

By 2030, NIST discourages organizations from using any cryptographic method that uses less than 112-bit security. However, from 2025 to 2030, NIST is expected to release documents on how to fully move to quantum cryptography. This raises the question: when should organizations find frameworks and have the necessary time to remove classical cryptography from their systems. International collaboration is needed to help mitigate attacks from quantum computers; the sharing of knowledge would make it easier to spot attacks. Based on the timeline below, NIST should, in 2028, have the United States government collaborating with other countries to create international standards for hybrid machines and cryptography. By 2035, organizations should begin transitioning from hybrid cryptography to fully quantum cryptography. 2035 should be the time organizations start transitioning to full quantum cryptography. The organizations had ten years to learn, understand, and work with quantum cryptography. Rather than removing classical algorithms with less than 112-bit security in 2030, it should start in 2035, so it gives users time to understand the technology and what it is capable of (Figure \ref{fig:pqcmain}).

\section{Conclusion}

Quantum cryptography is poised to revolutionize the cryptographic industry, but to achieve this, a hybrid cryptography approach must first be established. There are many ways to understand quantum cryptography using hybrid cryptography. First, there need to be formal publications of the Federal Information Processing Standards (FIPS) and Special Publications for hybrid cryptography. This allows users to understand what changes are happening and how to start the transition. The steps outlined above are the process that organizations undergo to transition from completely classical cryptography to hybrid cryptography. While this does not need to be done directly on the organization’s systems, there are now different ways to test the newer technology in a safe environment. This can be done using Open Quantum Safe and Qiskit. The Open Quantum Safe and Qiskit are excellent examples of how to gain hands-on experience with quantum encryption in a hybrid machine. Eventually, organizations will need to rely on hybrid machines to protect themselves against the risks associated with quantum cryptography. Some risks mentioned above include being vulnerable to data breaches and insecure installation of quantum cryptography. A good way to minimize these vulnerabilities is to start learning and testing hybrid cryptography now. Preventing the misuse of hybrid machines requires policymakers and lawmakers to create regulations, frameworks, and laws now. Creating guidelines on hybrid cryptography will allow for a smooth transition from classical to quantum cryptography. These creations can reduce the risks and hold those abusing hybrid cryptography responsible. There needs to be more urgency to move to hybrid cryptography now, so the transition to quantum cryptography is smooth. 

\printbibliography
\end{document}